\documentclass[11 pt]{article}

\usepackage[dvips]{graphicx}
\usepackage{amsmath}
\usepackage{latexsym}
\usepackage{amsthm}
\usepackage{amssymb}
\usepackage{color}
\usepackage[round,numbers,sort&compress]{natbib}

\title{\textbf{Effects of multiple occupancy and inter-particle interactions on selective transport through narrow channels: theory versus experiment}}

\author{Anton Zilman\thanks{ Corresponding author.  Address: LANL, POB 1663, MS B258, Los Alamos, N
M 87545, U.S.A., Tel.:~(505)-667-3216, e-mail: zilmana@lanl.gov}\\
Theoretical Biology and Biophysics Group\\ and Center for Non-Linear Studies\\
Theoretical Division, Los Alamos National Laboratory }

\date{}

\begin{document}

\maketitle
\abstract{ Many biological and artificial transport channels function without direct input of metabolic  energy during a transport event and without structural rearrangements involving transitions from a 'closed' to an 'open' state. Nevertheless, such channels are able to maintain efficient and selective transport. It has been proposed that attractive interactions between the transported molecules and the channel can increase the transport efficiency and that the selectivity of such channels can be  based on the strength of the interaction of  the specifically transported molecules with the channel. Herein, we study the transport through narrow channels in a framework of a general kinetic theory, which naturally incorporates multi-particle occupancy of the channel and non-single-file transport. We study how the transport efficiency and the probability of translocation through the channel are affected by inter-particle interactions in the confined space inside the channel, and establish conditions for selective transport. We compare the predictions of the model with the available experimental data - and find good semi-quantitative agreement. Finally, we discuss applications of the theory to the design of artificial nano-molecular sieves.
 }

\section{Introduction}

The proper functioning of living cells involves continuous transport of various molecules into and out of the cell, as well as between different cell compartments. Such transport requires  discrimination between different intra- and extra-cellular molecular signals and demands mechanisms for efficient, selective and specific transport \cite{alberts-book}. Specifically, transport devices must be able to selectively transport only certain molecular species while effectively filtering others, even very similar ones.

In certain cases, the selectivity and efficiency of the transport is achieved through direct input of metabolic energy during the transport event, in the form of the hydrolysis of ATP or GTP \cite{alberts-book}. However, in many cases, molecular transport is efficient and selective without the direct input of the metabolic energy and without large scale structural rearrangements that involve transitions from a 'closed' to an 'open' state during the transport event. Examples of transport  of this type include  the selective permeability of porins \cite{bezrukov-antibiotics-PNAS-2002,bezrukov-kullman-maltoporin-2000,schulten_glycerol,GlpF-science-2001,aquaporins-borgnia,aquaporins-book}, transport through the nuclear pore complex in eukaryotic cells \cite{mike_review,macara-review,stewart_review,lim-aebi-review,wente-review}, artificial nano-channels  and membranes, \cite{caspi-elbaum-2008,martin-antibody-science-2002,martin-DNA-2004,martin-apoenzymes-1997,tijana,akin-DNA-nanotubes-2007,polymer-nanotubes-2008} and other transport devices \cite{cytochrome-specificity-2000}. Ion channels \cite{jordan-review,hille-book,corry-chung-valence-selectivity-2006}, also belong to this class of transport devices - however the selectivity of ion channels depends on numerous factors that set them apart \cite{eisenberg-barriers-2007,corry-chung-valence-selectivity-2006} and place them beyond the scope of the present work.

Transport devices of this type  commonly contain a channel or a passageway through which the molecules translocate by facilitated diffusion. The selectivity and the efficiency of transport are usually based not merely on the molecule size but on a combination of the size, strength of the interaction with the channel, speed of the spatial diffusion through the channel, and channel geometry (cf. Figs. 1 and 2). \cite{bezrukov-antibiotics-PNAS-2002,lim-aebi-review,caspi-elbaum-2008,aquaporins-book,GlpF-science-2001,aquaporins-borgnia,bezrukov-kullman-maltoporin-2000,schulten_glycerol,mike_review,macara-review,wente-review,diffsuion-zeolites-review-2003,martin-DNA-2004,martin-apoenzymes-1997,martin-antibody-science-2002,berezhkovskii_ptr,noble-theory-1992,noble-theory-1991,stewart_review,polymer-nanotubes-2008,akin-DNA-nanotubes-2007}.   Moreover, a large body of experimental data shows  that the specifically transported molecules in many cases interact strongly with the channel (more strongly than the ones that are filtered out) and can transiently bind inside the channel \cite{tijana,akin-DNA-nanotubes-2007,mike_review,bezrukov-antibiotics-PNAS-2002,lim-aebi-review,GlpF-science-2001,schulten_glycerol,wente-review,bezrukov-kullman-maltoporin-2000,macara-review,martin-DNA-2004,martin-apoenzymes-1997,martin-antibody-science-2002,caspi-elbaum-2008,aquaporins-borgnia,aquaporins-book,stewart_review}.
Another important feature of such selective channels is that they are narrow, with a diameter comparable to the size of the transported molecules.

Understanding mechanisms of the selectivity of transport through such channels is an important biological question and also has important applications in nano-technology and nano-medicine. For instance, it impacts creation of artificial molecular nano-filters. In addition, it poses a fundamental physical question: how does one make a selective channel that is always open, and does not have a movable 'shutter' specifically attuned to its corresponding transported molecules? Another important goal is to establish to what extent the theoretical models capture the essential properties of transport through narrow channels by comparing the models to experimental data.

The precise mechanisms and the conditions for optimal selectivity of transport through such channels are still unknown. These systems span a wide spectrum of space and time scales and biological functions.
For instance, porins are involved in the transport of small molecules into and out of the cell. They typically have channel length of several nanometers and a diameter of a couple of nanometers, tuned to the size of their corresponding transported molecules (e.g. water or small sugars) \cite{aquaporins-book,aquaporins-borgnia,bezrukov-antibiotics-PNAS-2002,bezrukov-kullman-maltoporin-2000,GlpF-science-2001,schulten_glycerol}. The transport times through porins can be shorter than one millisecond \cite{bezrukov-antibiotics-PNAS-2002}. In another example, the nuclear pore complex regulates the transport between the cell nucleus and the cytoplasm. It has a diameter of approximately 30 nm and a length of 70 nm \cite{mike_review,wente-review,stewart_review,macara-review}. It  can pass molecules up to 30 nm in size, within transport times of several milliseconds \cite{musser_single,peters-1999-single-pore}. Artificial selective nano-channels  have been constructed several microns long and tens of nanometers in diameter that  selectively transport molecules of sizes in the range of tens of nanometers. Such artificial devices have been used to selectively transport various molecules: including molecular enantiomers, short DNA segments and synthetic polymers \cite{tijana,akin-DNA-nanotubes-2007,caspi-elbaum-2008,martin-apoenzymes-1997,martin-DNA-2004,martin-antibody-science-2002,polymer-nanotubes-2008}.

Nevertheless, it has been suggested that such channels might share common mechanisms of selectivity and efficiency. Recent theoretical works propose a mechanism of selectivity that relies on two crucial factors, transient trapping of the cargoes inside the pore and the resulting confinement of the cargoes in the limited space within the channel. In particular, by modeling the transport as diffusion in an effective potential, the authors of \cite{schulten_glycerol,berezhkovskii_ptr,strange-people-occusion-pnas-2006,we-NPC-plos-2007,bezrukov-asymmetric-2007,bezrukov-sites-2005,noble-theory-1991,noble-theory-1992} have  shown that the attractive interactions of the transported molecules with the channel, such as transient binding of the molecules to binding moieties, increase the transport efficiency. More precisely, without an attractive potential inside the channel, the particles entering the channel have a low probability of traversing it to the other side. Attractive interactions inside the channel slow down the passage and increase the \textit{probability}  of individual molecules to translocate through the channel \cite{we-NPC-plos-2007,gardiner-book,strange-people-occusion-pnas-2006,schulten_glycerol,bezrukov-sites-2005,schuss-eisenberg-1995,noble-theory-1991,noble-theory-1992,berezhkovskii_ptr}. This mechanism of transport enhancement has also been known as 'facilitated diffusion' in the field of membrane transport \cite{noble-theory-1991,noble-theory-1992,kussler,myoglobin-1966}.

However, space inside the channel is limited, and if  the molecules spend too much time inside the channel, they prevent entrance of new ones. The channel thus becomes jammed and the transport is diminished. To model the jamming of the channel, the authors of \cite{strange-people-occusion-pnas-2006,berezhkovskii-optimal-2005,kolomeisky-2006,bezrukov-sites-2005,bezrukov-asymmetric-2007} assumed that additional molecules cannot enter the channel already when one molecule is present inside. They showed that particles whose interaction with the channel is weaker than the optimal, have a low probability of traversing the channel, while particles that interact too strongly with the channel jam the transport.  This allows discrimination between the molecules based on the strength of their interaction with the channel, and provides a mechanism of selective transport; transmission efficiency is optimized for a particular interaction strength and rate of transport. Optimal trapping time, which maximizes the transmitted current, has also been demonstrated for single-file transport in \cite{chou-PRL-single-file-1998}.

However, during transport, the channel can be occupied by multiple molecules, which cannot bypass each other, or do so only in the limited fashion, due to the confinement in the limited space inside the channel \cite{berezhkovskii-hummer,chou-PRL-single-file-1998,schutz-review-2005,chou-tasep-2003,chou-zeolites-PRL-1999edp,derrida-mukamel-exclusion-1992,schulten_glycerol}. The transport is not necessarily single-file: the number of molecules that can be present at a position along the channel depends on the ratio of the channel diameter to the molecule size. We must also recognize that the transport properties of narrow channels are not dominated by the equilibrium thermodynamic channel-cargo interactions \textit{per se}, but by the  rates at which the cargoes enter, translocate through and exit from the  channel with a potentially  complicated geometry \cite{alberts-book,mike_review,diffsuion-zeolites-review-2003,bezrukov-sites-2005,chou-PRL-single-file-1998,bezrukov-asymmetric-2007,we-NPC-plos-2007,aquaporins-book,entropic-transport-PRL-2006}. For instance, the trapping time in the channel can be limited by the time it takes to find a narrow exit from the channel by diffusion. This phenomenon is known as entropic trapping \cite{diffsuion-zeolites-review-2003,entropic-transport-PRL-2006,chou-zeolites-PRL-1999edp}. In the case when the rates are determined solely by the interaction strength, stronger interactions with the channel imply slower rates and higher trapping times (cf. Figs. 1 and 2)\cite{berezhkovskii-optimal-2005,berezhkovskii_ptr,kolomeisky-2006,strange-people-occusion-pnas-2006,schulten_glycerol,gardiner-book}.

Understanding the effects of multiple channel occupancy and jamming on the transport selectivity is especially pertinent to the analysis of single molecule tracking experiments  \cite{musser_single,kubitschek_single,yang-musser-JCB-2006} and the design of artificial nano-molecular filters \cite{caspi-elbaum-2008,martin-DNA-2004,martin-antibody-science-2002,tijana,martin-apoenzymes-1997,polymer-nanotubes-2008,akin-DNA-nanotubes-2007}.

In this paper, we analyze transport through narrow channels in a framework of a general kinetic model based on exclusion process theory as a function of the kinetic parameters of transport Specifically, we examine the rates of entrance, hopping through and exit from the channel. We extend the previous work to include multiple occupancy and inter-particle interactions inside the channel beyond single file.
We investigate how the concentration of the cargoes, the channel length and radius, the dimensions of the transported molecules and the interactions between them inside the channel influence the transport.
An important goal of this paper is to explore whether a theory that has only two essential ingredients: 1) transient trapping of the molecules inside the channel and 2) inter-particle crowding due to the confinement in the limited space inside the channel, can provide an adequate explanation of the selective transport through narrow channels by comparing the predictions of the theory with the available experimental data.

The paper is organized as follows. We first discuss a channel that consists only of one 'site' and then two-sites. Next, we discuss  transport in a uniform symmetric channel of arbitrary length, for both single-file and non-single-file transport, and establish conditions for optimal transport. We then discuss the transition between two transport regimes, jammed and un-jammed, and establish the relative contribution of the jamming of the channel entrance as compared to crowding inside the channel, to the transport selectivity and efficiency. Next, we compare predictions of the theory with the experimental data. We conclude with discussion of the results, their relation to the previous work, and consider potential applications.

\section{Inter-particle interactions inside the channel }
A transport channel can be represented as a chain of positions (sites), as illustrated in Figs. 1 and 2 \cite{berezhkovskii-hummer,gardiner-book,kolomeisky-2006,we-NPC-plos-2007,bezrukov-sites-2005,chou-PRL-single-file-1998,chou-zeolites-PRL-1999edp,schutz-review-2005,derrida-mukamel-exclusion-1992,schulten_glycerol}.
The particles attempt to enter the channel at a given position, with an average rate $J$ and subsequently hop back and forth between adjacent sites, if those are not fully occupied, until they either reach the rightmost or leftmost sites, from where they can hop out of the channel. Hopping out from the rightmost site represents the particle reaching its destination compartment, while hopping out from  the leftmost site channel represents an abortive transport event, where the molecule does not reach its destination (cf. Figs. 1 and 2). In the continuum limit, when the distance between the adjacent sites tends towards zero (and their number to infinity), with an appropriate choice of the transition rates between the sites, the problem can be reduced to diffusion in an effective continuous potential \cite{we-NPC-plos-2007,bezrukov-sites-2005,gardiner-book,shuss-eisnberg-microscopi-2001} (cf. also Appendix). Note that the discrete positions (sites) do not represent the actual binding sites inside the channel. Rather, they are a convenient computational tool that allows one to explicitly take into account competition for space and interactions between multiple particles inside the channel \cite{gardiner-book,we-NPC-plos-2007,bezrukov-sites-2005,berezhkovskii-hummer,chou-PRL-single-file-1998,chou-zeolites-PRL-1999edp,derrida-mukamel-exclusion-1992}. The distance between the positions reflects the size of the particles.

As the particles accumulate in the limited space inside the channel, they start to interfere with the movement of the neighboring particles and prevent the entrance of new ones.  We must differentiate between the speed, the efficiency, and the probability of transport. The \textit{speed} is determined by the time the particles spend in the channel. The \textit{efficiency} of transport is determined by the fraction of the impinging flux that reaches the rightmost end. It depends on the kinetic parameters of the channel, such as transition rates inside the channel and the exit rates at its ends. The \textit{selectivity} of transport is determined by the different efficiencies at different values of the kinetic parameters  \cite{berezhkovskii-optimal-2005,berezhkovskii_ptr,schuss-eisenberg-1995,we-NPC-plos-2007,bezrukov-asymmetric-2007,bezrukov-sites-2005,chou-PRL-single-file-1998}. Transport efficiency is different from the \emph{probability} that an individual particle translocates through the channel. The latter  is  defined as the fraction of the particles that reach the exit \emph{after} entering the channel. We discuss these issues in detail below.
\subsection{One site channel}\label{section-single-site}
To get started, let us consider a 'one-site' channel \cite{bezrukov-sites-2005,gardiner-book}, where all the internal spatial and energetic structure of the channel is absorbed into the forward and the backward exit rates $r_{\rightarrow} $, $r_{\leftarrow} $.

Kinetic diagram of such a 'one-site' channel is shown in Fig. 1\textbf{B}. The state of the channel is specified by the particle  density ($0\leq n\leq 1 $) at the channel site (or, in other words, the probability of the channel to be occupied). It  obeys the following kinetic equation \cite{gardiner-book,bezrukov-sites-2005}:
\begin{equation}
\dot{n}=J(1-n)-(r_{\leftarrow}+r_{\rightarrow})n
\end{equation}
which takes into account that the particles can enter the channel only if it is not occupied. The average time a particle spends inside the channel is $\tau=1/(r_{\leftarrow}+r_{\rightarrow})$
\cite{bezrukov-sites-2005,gardiner-book}.

At steady state ($\dot{n}=0 $) we get  for the average density and the forward flux:
\begin{eqnarray}\label{one-site-equation}
n&=&\frac{J}{J+r_{\leftarrow}+r_{\rightarrow}}\\
J_{\text{out}}&=&nr_{\rightarrow }=\frac{Jr_{\rightarrow}}{J+r_{\leftarrow}+r_{\rightarrow}}=\frac{Jr_{\rightarrow}}{J+1/\tau}\nonumber
\end{eqnarray}
As mentioned above, we define the transport efficiency as the ratio of the transmitted flux to the entering flux, $\text{Eff}=J_{\text{out}}/J $. Thus, from the equation (\ref{one-site-equation}) we learn  that the transport efficiency $\text{Eff}(J,r_{\leftarrow},r_{\rightarrow})= \frac{r_{\rightarrow}}{J+r_{\leftarrow}+r_{\rightarrow}} $  is a monotonic function of both the forward exit rate $r_{\rightarrow} $ and the backward exit rate $r_{\leftarrow}$. Therefore, for the 'one-site channel' there is no optimal combination of the exit rates that would  maximize the transport. As we shall see, this is not the case for longer channels. However, even a single site channel can have more interesting behavior, if the forward and the backward exit rates are not independent \cite{bezrukov-sites-2005,bezrukov-asymmetric-2007} (cf. Appendix).
\subsection{Two site channel}
Let us consider now a longer channel consisting of two sites: 1 and 2. This is the shortest channel that explicitly takes into account the asymmetry between the channel entrance and exit, and exhibits non-trivial transport properties \cite{schulten_glycerol,bezrukov-sites-2005,strange-people-occusion-pnas-2006,chou-zeolites-PRL-1999edp,bezrukov-asymmetric-2007}. The kinetics of transport through such a channel is illustrated in Fig. 1 \textbf{C}  and \textbf{D}. The particles enter the channel at the entrance site 1 with an average flux $J $, if it is un-occupied. The backward exit  rate to the left from  site 1 is $r_{\leftarrow }$ and the forward exit  rate to the right from site 2, is $r_{\rightarrow} $. Once inside, a particle can hop back and forth between sites 1 and 2 with rates $r_{12}$ and $r_{21}$, respectively, if the target site is un-occupied. The exit rates $r_{\rightarrow} $ and $r_{\leftarrow} $ can be thought of as the 'off' rates for the release of the particles from the channel \cite{bezrukov-sites-2005}. For simplicity, in this section we assume that the channel is internally uniform and symmetric with $r_{12}=r_{21}=r$ and $r_{\rightarrow}=r_{\leftarrow}=r_o $ and that each site can be occupied only by one particle.

The state of the channel is characterized by the average occupancies of the sites, $0\leq n_1\leq 1$ and $0\leq n_2\leq 1 $. For an internally uniform channel, these average occupancies can also be viewed  as the probabilities that the sites $1$ and $2$ are occupied by a particle \cite{chou-tasep-2003,schutz-review-2005,schulten_glycerol}. The kinetic equations describing transport through such a channel are (Fig. 1\textbf{C})
\begin{eqnarray}\label{two-site-equation}
\dot{n_1}&=&J(1-n_1)-r_on_1+rn_2(1-n_1)-rn_1(1-n_2)\\
\dot{n_2}&=&rn_1(1-n_2)-rn_2(1-n_1)-r_on_2\nonumber
\end{eqnarray}
and the transmitted flux is $J_{\text{out}}=r_on_2$.

The transport efficiency $\text{Eff}(r_o)=J_{\text{out}}/J$ is the fraction of the flux $J$ that  exits the channel to the right. Solving equations (\ref{two-site-equation}) at steady state ($\dot{n}_1=\dot{n}_2=0)$, one gets for the transport efficiency:
\begin{equation}\label{eq-efficiency-two-sites}
\text{Eff}(r_o)=J_{\text{out}}/J= \frac{rr_o}{r_o(2r+r_o)+J(r+r_o)}
\end{equation}

Importantly, unlike in the one-site case, for a given entrance flux $J$, the transport efficiency $\text{Eff}(r_o)$ has a maximum at a certain value of the exit rate $r_o^{\text{max}}=\sqrt{Jr}$. This provides a mechanism of selectivity; only particles whose residence time in the channel (determined by the interactions of the particles with the channel) is close to $1/r_o^{\text{max}} $ are transmitted efficiently \cite{we-NPC-plos-2007,berezhkovskii-optimal-2005,strange-people-occusion-pnas-2006,bezrukov-sites-2005,kolomeisky-2006,chou-PRL-single-file-1998}.

The total efficiency $\text{Eff}=J_{\text{out}}/J $ is influenced by two different effects, the jamming of the channel entrance and the mutual interference between the particles \textit{inside} the channel. The flux that actually enters the channel is $J_{\text{in}}=J(1-n_1)$. The remaining portion of the flux, $Jn_1$, does not enter the channel because the entrance site $1$ is occupied $n_1$ fraction of the time. The fraction of the entering current $J_{\text{in}}$ that reaches the exit on the right determines the transport \emph{probability} $P_{\rightarrow}=J_{\text{out}}/J_{\text{in}}$, which characterizes transport \emph{through} the channel. From the equations (\ref{two-site-equation})
\begin{equation}
P_{\rightarrow}=\frac{r}{2r+r_o}
\end{equation}

Very importantly, $P_{\rightarrow}$ is independent of the flux $J$ and is equal to the efficiency in the single-particle limit, $J\rightarrow 0$. That is, it is equivalent to the probability of a single particle to translocate through the channel
when no other particles are present.  Thus, surprisingly, the crowding of the particles inside the channel \emph{does not}, on average, influence their movement \emph{through} the channel. We discuss this effect at length below.

To summarize this section, selective transport can arise from a balance between two competing effects, enhancement of the transport by the transient trapping and the eventual jamming of the channel if the trapping times are too high\cite{bezrukov-asymmetric-2007,bezrukov-sites-2005,berezhkovskii-optimal-2005,chou-PRL-single-file-1998,we-NPC-plos-2007}.

\subsection{Channel of arbitrary number of sites}\label{section-N-sites-with-exclusion}

In this section we study transport through a channel of arbitrary length, which is modeled as a chain of  $N$  positions (sites): \textit{1, 2...i...N}. Particles enter  at site $M$ (not necessarily the leftmost one) with an average flux $J$ if the entrance site is not fully occupied. Once inside the channel, a particle at  site $i$  can hop to an adjacent site $i\pm 1 $ if the latter is not fully occupied. In order to model the excluded volume interaction  between the particles in the channel (they can bypass each other only in a limited fashion), we introduce the maximal site occupancy $n_m$, which depends on the ratio of the channel diameter to the size of the particles.  When at an outermost site $1$ or $N$, a particle can leave the channel with the rate $r_{\leftarrow}$ and $r_{\rightarrow}$, respectively, or hop into the channel with the average rate $r_{1\rightarrow 2}$ or $r_{N\rightarrow N-1}$, respectively.  The kinetics of this process is illustrated in Fig. 2 \textbf{A}. The rates $r_{i\rightarrow i\pm1}$ determine the speed with which the particles diffuse through the channel, while the exit rates $r_{\leftarrow}$ and $r_{\rightarrow}$ reflect how fast the particles can leave the channel.  The kinetic rates $r_{i\rightarrow i\pm1}$, $r_{\leftarrow}$ and $r_{\rightarrow}$ are determined by the microscopic interactions of the particles with the channel, and by its geometry. As before, the exit rates $r_{\rightarrow} $ and $r_{\leftarrow} $ can be thought of as the 'off' rates for the release of the particles from the channel \cite{bezrukov-sites-2005}. In general, with a proper choice of the transition rates, $r_{i\rightarrow i\pm 1}/r_{i\rightarrow i\pm 1}=\exp(U_{i+1}-U_{i})/2)$ in the continuum limit the model reduces to diffusion in the potential $U(x)$ \cite{gardiner-book,bezrukov-sites-2005,shuss-eisnberg-microscopi-2001,berg-book}.
Trapping of the particles in the channel corresponds to low exit rates $r_{\rightarrow},r_{\leftarrow}<r$ \cite{gardiner-book,bezrukov-asymmetric-2007,bezrukov-sites-2005,redner-book}.

For simplicity, we assume that the channel is internally uniform, such that all the internal transition rates are equal,
$r_{i\rightarrow i\pm 1}=r$ for all $i$. At any time $t$, the state of the channel is specified by the number densities of the particles at each site $n_1, n_2,...,n_i,...n_{N} $. The kinetics of transport through such a channel is  described by the following equations \cite{schulten_glycerol,schutz-review-2005,chou-zeolites-PRL-1999edp,derrida-mukamel-exclusion-1992,chou-PRL-single-file-1998,gardiner-book}

\begin{eqnarray}\label{kinetics_with_exclusion}
\dot{n_i}&=&J\delta_{i,M}(1-\frac{n_i}{n_m})+rn_{i-1}(1-\frac{n_i}{n_m})+rn_{i+1}(1-\frac{n_i}{n_m})-rn_i(1-\frac{n_{i-1}}{n_m})-rn_i(1-\frac{n_{i+1}}{n_m})\nonumber\\
&=&J\delta_{i,M}(1-\frac{n_i}{n_m})+r(n_{i-1}+n_{i+1}-2n_i)
\end{eqnarray}
with the boundary conditions at sites $1$ and $N$
\begin{eqnarray}\label{kinetics_with_exclusion-2}
&&\dot{n_1}=J\delta_{1,M}(1-\frac{n_1}{n_m})-r_{\leftarrow}n_1-rn_1(1-\frac{n_2}{n_m})+rn_2(1-\frac{n_1}{n_m})\nonumber\\
&&=J\delta_{1,M}(1-\frac{n_1}{n_m})-(r+r_{\leftarrow})n_1+rn_2\\
&&\dot{n}_{N}=-r_{\rightarrow}n_{N}-rn_{N}(1-\frac{n_{N-1}}{n_m})+rn_{N-1}(1-\frac{n_{N}}{n_m})=-(r+r_{\rightarrow})n_{N}+rn_{N-1}\nonumber
\end{eqnarray}
where the $\delta$-function is $\delta_{i,j}=1 \;\;\text{if}\;\; i=j$ and zero otherwise.
The terms $n_i(1-n_{i\pm1}/n_m) $ in  equations (\ref{kinetics_with_exclusion}) and (\ref{kinetics_with_exclusion-2}) reflect the fact that a particle can jump to the next site only if it is not fully occupied, $n_{i\pm1}<n_m$. Importantly, for an internally uniform channel, at all the internal sites the cross-terms of the form $n_in_{i\pm1}$ cancel out \cite{schutz-review-2005,chou-PRL-single-file-1998}. For such uniform channels, the equations (\ref{kinetics_with_exclusion}) and (\ref{kinetics_with_exclusion-2}) are exact and $n_i/n_m$ is equivalent to the probability of  a site $i$ to be occupied \cite{schutz-review-2005,chou-PRL-single-file-1998}.
Obstruction of the space inside the channel by the particles inside of it affects only the \textit{entrance} to the channel at site $M$.

We define the \emph{efficiency} of transport as the ratio of the forward  exit current $J_{\text{out}}=r_{\rightarrow}n_{N}$ to the incoming flux $J$, $\text{Eff}(r_o)=J_{\text{out}}/J$. It is the fraction of the incoming flux that traverses the channel. Note that the efficiency is different from  the probability of individual particles to traverse the channel \emph{after} they have entered, because some of the particles attempt to enter the channel and are rejected if the entrance site is occupied.

The linear equations (\ref{kinetics_with_exclusion}),(\ref{kinetics_with_exclusion-2}) can be solved analytically for any $N$ \cite{chou-PRL-single-file-1998}.
In the fully symmetric case, when the forward and the backward exit rates from sites $1$ and $N$ are equal, $r_{\leftarrow }=r_{\rightarrow }=r_o $, the  efficiency is given by:
\begin{eqnarray}\label{Jout_with_exclusion}
\text{Eff}(r_o)=\frac{(r+(M-1)r_o) r_o}{r_o(2r+(N-1)r_o)+\frac{J}{n_m}(r+(N-1)r_o+(M-1)(N-M)r_o^2/r)}
\end{eqnarray}
(for $M<N/2 $).

Note that in the single particle diffusion limit, $J\rightarrow 0 $,  the efficiency $\text{Eff}\rightarrow M/(N+1)$ without trapping ($r_o\rightarrow r$) and $\text{Eff}\rightarrow 1/2$  for strong trapping ($r_o\rightarrow 0$), in accord with  known results \cite{bezrukov-sites-2005,berezhkovskii_ptr,we-NPC-plos-2007,gardiner-book,peters-1999-single-pore,berg-book,chou-PRL-single-file-1998}. Essentially, without transient trapping, the probability to traverse the channel is low and reaches one half for very strong trapping.

One may rewrite equation (\ref{Jout_with_exclusion}) in terms of the trapping time, which is equal to $\tau=\frac{N}{2r_o} $ \cite{pearson-BJ-2005,redner-book,bezrukov-sites-2005} to arrive at
\begin{eqnarray}
\text{Eff}(r_o)=\frac{(2\tau r+(M-1)N)N}{(N-1)N+4\tau r+\frac{J'}{n_{\text{m}}}(\frac{4}{N}(\tau r)^2+(N-1)\tau r+(M-1)(N-M))}
\end{eqnarray}
where $J'=J/r $ is the normalized flux. Note that the transport efficiency does not depend on the absolute values of the transport rates $r$ and $r_o$, but only on the normalized parameters $\tau r$ and $J/r$. This means that the transport efficiency can be the same for different particles, even if the kinetics of their transport through the channel is very different from each other, as long as they possess the same $\tau r$ and $J/r$.

As already seen in the two-site case, the transport efficiency $\text{Eff}(r_o)$ of equation (\ref{Jout_with_exclusion}) has a maximum at a certain value of the exit rate (for $M=1$) of
\begin{equation}\label{eq-optimal-rate}
r_o^{\text{max}}/r=\sqrt{\frac{J/(rn_{\text{m}})}{N-1}}
\end{equation}
and the maximal flux at this rate is
\begin{equation}\label{eq-optimal-current}
J_{\text{max}}=\frac{J}{\left((N-1)r_o^{\text{max}}/r+1\right)^2+1}
\end{equation}
(cf. Appendix for $M\neq 1$).

This feature provides a mechanism of selectivity; only the particles whose exit rate is close to  the optimal one, $r_o^{\text{max}}$, are transmitted efficiently. Particles with exit rates higher than the optimal have a higher chance of returning back because they don't spend enough time inside the channel in order to reach the farther exit on the right side. On the other hand, due to the limited space inside the channel, the particles with the exit rates lower than optimal spend so much time in the channel that it gets jammed and the entrance of new particles is inhibited \cite{berezhkovskii-optimal-2005,we-NPC-plos-2007,strange-people-occusion-pnas-2006,kolomeisky-2006,schulten_glycerol,bezrukov-asymmetric-2007}.

Equations (\ref{Jout_with_exclusion}) and (\ref{eq-optimal-rate}) qualitatively agree with the  results of \cite{kolomeisky-2006,strange-people-occusion-pnas-2006,berezhkovskii-optimal-2005,bezrukov-asymmetric-2007}, which assumed that only one molecule can occupy the channel. Figure 3 shows how the transport depends on the channel length $N$, the entrance flux $J$, the exit rate $r_o$ and the effective channel width $n_m$. Note that the optimal exit rate $r_o$ decreases with the channel length $N$; for longer channels, a particle has to spend more time in the channel in order to reach the other end.  Also note that the optimal rate of equation (\ref{eq-optimal-rate}), $r_o^{\text{max}}/r$, is less than one for $J/r<N-1$; the optimal interaction is attractive for small currents and long channels.  We elaborate on this issue in Appendix.

\subsection{Transport efficiency vs. translocation probability}
In this section, we elaborate on why the flux through the channel decreases in the limit of very low exit rates? Is this because new particles cannot enter or because the particles inside the channel interfere with each other's passage?

The fraction of the incoming flux $J$ that actually enters the channel is $J_{\text{in}}=J(1-n_1/n_m)$. The remaining portion of the flux $J\frac{n_1}{n_m}$ cannot enter because the entrance site is occupied on average $\frac{n_1}{n_m}$ fraction of the time (cf. Sec. 4 for calculation of the densities). The total efficiency is determined by two quantities: \emph{i)} the fraction of the flux  that enters the channel $J_{\text{in}}$ and \emph{ii)} the fraction of the particles that upon entering the channel, actually reach the rightmost end. The latter defines the \emph{probability} $P_{\rightarrow }=J_{\text{out}}/J_{\text{in}}$ of a particle exiting to the right \textit{after it has entered} the channel and is given by
\begin{equation}
P_{\rightarrow}=\frac{J_{\text{out}}}{J_{\text{in}}}= \frac{r+(M-1)r_o }{2r+(N-1)r_o}
\end{equation}
Remarkably, it is independent of the flux $J$ and is \textit{exactly} equal to the efficiency in the single particle transport limit, $J\rightarrow 0$. This means that in unform channels the interactions between the particles in the channel \emph{do not} affect the transport probabilities of the individual particles. The effect of the channel occupancy manifests only in the jamming at the entrance site.

\section{Optimal transport and jamming}

The lower the exit rate $r_o$, the longer the time that the particles spend inside the channel. The trapping  time varies as $\tau=\frac{N}{2r_o}$ \cite{pearson-BJ-2005,redner-book,bezrukov-sites-2005}. As shown in the previous section, at very small exit rates $r_o$, the trapping time is so high that the channel becomes jammed. Thus, the transport efficiency is maximized at the particular exit rate $r_o^{\text{max}}$. Inspection of the Fig. 3\textbf{A} reveals two distinct transport regimes, roughly separated by the maximum of the transport efficiency at $r_o=r_o^{\text{max}}$. At the high values of $r_o>r_o^{\text{max}}$ the transport of individual particles is essentially unhindered by the presence of the others, as evidenced by the fact that the transport efficiency curves collapse onto the dashed line, representing the zero-current, single-particle limit (Fig. 3\textbf{A}).
At the low values of the exit rate where $r_o<r_o^{\text{max}}$, the accumulating  particles start to obstruct the entrance of the new ones. This feature provides a natural definition for the 'jamming transition' around the $r_o=r_o^{\text{max}} $.

Solving equations (\ref{kinetics_with_exclusion}) and (\ref{kinetics_with_exclusion-2}), we get for the density profile of the particles inside the channel, at the steady state:
\begin{eqnarray}\label{equation-ni}
n_i=\frac{J(r+(N-i)r_o)}{r_o(2r+(N-1)r_o)+J(r+(N-1)r_o)}
\end{eqnarray}
(for $M=1$, $n_m=1$). Note that unlike the equilibrium distribution, the maximum of the density profile is near the channel entrance at site $1$.

The total number of the particles in the channel is
\begin{eqnarray}\label{eq-Ntot-jam}
N_{\text{tot}}=\sum_{i=1}^{N}n_i=\frac{N}{2}\frac{J(2r+(N-1) r_o)}{r_o(2r+(N-1)r_o)+J(r+(N-1)r_o)}
\end{eqnarray}
Note that in the limit $r_o\rightarrow 0$, $N_{\text{tot}}\rightarrow N$. That is, the particles accumulate and never leave the channel. Therefore, from equation (\ref{eq-Ntot-jam}) one finds that at the point of the jamming transition, $r_o=r_o^{\text{max}} $, the number of the particles in the channel is
\begin{equation}\label{equation-Nmax}
N_{\text{jam}}=N\frac{J/r+2\sqrt{\frac{J/r}{N-1}}}{2\left(J/r+2(\sqrt{\frac{J/r}{N-1}}+\frac{1}{N-1})\right)}
\end{equation}
Equation (\ref{equation-Nmax}) has important consequences (cf. Fig. 4 for illustration). It shows that for long channels, where  $N\gg J/r $, the fraction of the occupied sites at the jamming transition tends to one half: $N_{\text{jam}}/N\rightarrow 1/2 $ when $N\gg J/r $. This means that long channels can be filled up almost to half of their maximal capacity $N$ before the jamming effects start to matter. For the occupancies  below the jamming transition, the particles travel through the channel essentially unhindered. These effects are illustrated in Fig. 3\textbf{A} and Fig. 4\textbf{A}. This might explain why experiments on transport through narrow channels often measure apparent diffusion coefficients that are almost as large as those for the free diffusion \cite{ribbeck-gorlich-EMBO-2001,peters-1999-single-pore}.

\subsection{Jamming and saturation of the flux through the channel}
Although the transport efficiency $\text{Eff}(r_o,J)$ decreases with the increasing flux $J$, the total transmitted flux $J_{\text{out}}=J\;\text{Eff}(r_o,J) $ saturates at large fluxes ($J/r\rightarrow \infty $) to the limiting value
\begin{equation}
J_{\text{out}}^{\infty}/r=n_m\frac{r_o/r}{1+(N-1)r_o/r}
\end{equation}

(for $M=1$). This saturation of the transmitted flux at large incoming flux $J$ is another manifestation of the jamming of the channel entrance by the particles inside. Indeed, equation (\ref{equation-ni}) shows that the density at the entrance $n_1$ tends to $n_1=1$, as $J/r\rightarrow \infty $. In other words, the flux saturates because no more particles can enter the channel.  This is neatly summarized by the observation that $n_1=J_{\text{out}}/J_{\text{out}}^{\infty}$.

By contrast, the exit site $N$ is not completely blocked even at high
$J$ and $n_N\rightarrow 1/(1+(N-1)r_o/r $) as $J\rightarrow\infty$. Thus, even at very large fluxes, when the entrance site is completely blocked, the channel is not fully occupied. From equation (\ref{equation-Nmax}), the number of particles in the channel is $$N_{\text{tot}}^{\infty}=\frac{N}{2}\frac{2r+(N-1)r_o}{r+(N-1)r_o} $$
In particular, for long channels ($N-1\gg r/r_o$), the channel occupancy in the saturated limit is $N_{\text{tot}}/N=1/2$. Also note that the saturated flux is proportional to $n_m$, and that it decreases with $r_o/r$.

The results of this section closely parallel Michaelis-Maenten kinetics of multi-step enzymatic reactions \cite{nemenman-sinitsyn} and are important for the estimation of binding affinities from the channel transport experiments \cite{ribbeck-gorlich-EMBO-2001,kopito-elbaum-PNAS-2007,timney-mike-JCB-2006}, as well as for comparison with experiments on flux through artificial nano-channels ( - Sec 4). \ref{section-comparison-with-experiment}.


\section{Comparison with experiments}\label{section-comparison-with-experiment}
In experimental systems, the exit rates and the rates of transport through the channel are determined by a potentially complicated kinetics of binding and unbinding  inside the channel. Can the theory adequately describe these experiments? Facilitated diffusion theories produced results consistent with the experimental observations of the transport of gases through functionalized membranes \cite{noble-theory-1991,noble-theory-1992,kussler} enhancement of transport of oxygen by myoglobin \cite{myoglobin-1966} and the transport through bacterial porins\cite{schulten_glycerol}. In this section, we compare the theoretical predictions of this paper with the  experiments of Ref. \cite{martin-DNA-2004}.

Briefly, in the experiments of Ref.\cite{martin-DNA-2004} that we chose for comparison with theoretical predictions, transport of short DNA segments through artificial nano-channels was studied. The flux of the DNA segments through these channels was measured in two cases: 1) empty channels and 2) channels were lined with single-stranded DNA hairpins, grafted to the walls. Each hairpin  has a stretch of 18 un-paired bases in the middle. The transported particles were 18 base ssDNA segments with the sequence complementary to the un-paired regions of the ssDNA hairpins inside the channels. Thus, the transported DNA segments can transiently hybridize with the DNA grafted inside the channel. That investigation found that the flux through the DNA-containing channels is higher than through the channels without DNA hairpins inside,  providing evidence that the transient trapping indeed facilitates transport through nano-channels. However, eventually the interactions between the particles in the limited space inside the nano-channel  block the passage, causing the transmitted flux to saturate with the increase in the incoming flux. This is another signature of the transient trapping discussed above  in Sec. 3 \cite{martin-DNA-2004,martin-antibody-science-2002,martin-apoenzymes-1997,schulten_glycerol}.

The radius of an empty channel is $R\simeq 6$ nm  and the channel length is $L=6\mu$\cite{martin-DNA-2004}. The grafted ssDNA hairpins reduce the passageway radius, which,  for the purposes of comparison with the theory, we roughly estimate as $R\simeq 3$ nm for the channels with DNA grafted inside. Using the value for the gyration radius of the transported DNA segments $S\simeq 1$ nm \cite{martin-DNA-2004,doibook}, we estimate $n_m=6$ for the empty channels and $n_m=3$ for the channels with the grafted DNA hairpins inside. Furthermore, we estimate the incoming flux as $4D_{\text{out}}c R$ \cite{berg-book}, where $D_{\text{out}}=\frac{k_BT}{6\pi\eta S_H} $ is the diffusion coefficient of the transported DNA coils outside the channel \cite{doibook}; $\eta$ is the viscosity of the solvent, $S_H\simeq 0.7 S$ is the hydrodynamic radius of the coils \cite{doibook} and  $c$ is the outside concentration of the transported DNA \cite{gardiner-book,berg-book}. To model the finite capacity of the channel, we estimate the number of available positions in the channel as $N=L/(2S)$, where $L=6\mu$ \cite{berezhkovskii-hummer,chou-PRL-single-file-1998,chou-zeolites-PRL-1999edp,schulten_glycerol}(cf. also Appendix).

Finally, $r_o/r=\frac{4}{\pi}\frac{D_{\text{out}}}{D_{\text{in}}}\frac{L}{N R}Z$, where $Z$ is the reduction in the exit rate due to the trapping inside the channel \cite{bezrukov-sites-2005}(cf. also Appendix). We return to the question of how $Z$ is related to the actual binding energy below. The ratio $D_{\text{out}}/D_{\text{in}}$ and $Z$ are the two independent fitting parameters of the model (note that the $r$ and $r_o$ appear as independent parameters in eq.(\ref{Jout_with_exclusion}).

We first tested the model for the case without DNA segments attached inside the channel. The data (black dots) and the fit (black line) with $D_{\text{in}}/D_{\text{out}}=0.42$ and $Z=1$ are shown in black dots in Fig. 5\textbf{A}.
Analogously, for the channels with the DNA hairpins inside, the fit of the equation (\ref{Jout_with_exclusion}) to the data (red dots) is shown in the red line in Fig. 5\textbf{A}, with the best fitting parameters $D_{\text{in}}/D_{\text{out}}=0.004$ and $Z=0.00007$. Note that the diffusion coefficient is significantly reduced in the narrow channel filled with the grafted DNA hairpins \cite{peters-1999-single-pore}.

As expected, the transient binding of the transported DNA segments to the DNA hairpins inside reduces the exit rate $r_o$ by a factor $Z=0.00007$. Because this energy is influenced by many factors that are poorly understood \cite{martin-DNA-2004}, one can not easily connect the value of Z to the actual binding energy between the transported DNA and DNA hairpins. However, if the reduction in the exit rate is indeed determined mainly by the effective binding energy $\epsilon$, then the function  $Z\sim \exp(-\epsilon/kT)$ should describe the \emph{trend} in the dependence of $Z$ on  $\epsilon$ \cite{gardiner-book,bezrukov-sites-2005,kolomeisky-2006,chou-PRL-single-file-1998,schulten_glycerol}.  The authors of \cite{martin-DNA-2004} measured fluxes through the channel for DNA segments possessing different numbers of mismatches to the DNA grafted inside and found that the flux decreases with the number of mismatches. Thus, assuming as a first approximation that the binding energy $\epsilon $ decreases linearly with the number of mismatches $n$, so that $\epsilon (n)=\epsilon_{n=0}(18-n)/18 $ we get $Z=\exp(\ln(0.00007)(n-18)/18)$. The prediction of equation (\ref{Jout_with_exclusion}) with this choice of $Z$ is compared with the data in Fig. 5 \textbf{B}. It shows that this simple estimate correctly reproduces the trend in reduction of the transmitted flux with the number of mismatches. Note that there are no additional fitting parameters used in this figure.

That the simplified theory developed in this paper can correctly reproduce the trends in the observed fluxes, and even gives semi-quantitative fit of the data for reasonable values of the parameters, is encouraging. This demonstrates that a theory that is built upon only two essential assumptions, 1) facilitation of diffusion by the transient trapping inside the channel and 2) mutual interference between the particles crowded in the confined space inside the channel, does provide an adequate explanation of the experimental data.  Moreover, the theory provides verifiable predictions about how the flux should change with the channel diameter and length, as well as the particle size and concentration, as described in section 3. Comparison of these theoretical predictions with future quantitative experiments will lead to further ramification of the theoretical approach and will facilitate the design of artificial selective nano-channels with desired properties.

Discussion of other effects observed in \cite{martin-DNA-2004}, which are attributable to a conformational transition of the hairpin layer during transport, is outside the scope of the present work.

\section{Summary and discussion}
Proper functioning of living cells requires constant transport of different molecular signals into and out of the cell, as well as between different cell compartments. To carry out this task, the living cells have evolved various mechanisms for efficient and selective transport.

One class of transport devices comprises narrow channels whose diameter is comparable to the size of the molecules transported through it. Examples include selective transport through the nuclear pore complex, bacterial porins, \cite{stewart_review,aquaporins-borgnia,aquaporins-book,alberts-book,schulten_glycerol,bezrukov-kullman-maltoporin-2000,mike_review,lim-aebi-review,wente-review,bezrukov-antibiotics-PNAS-2002,GlpF-science-2001,macara-review} and other non-biological transport systems such as zeolites \cite{entropic-transport-PRL-2006,chou-zeolites-PRL-1999edp,diffsuion-zeolites-review-2003}.
A crucial feature of such transport channels is their ability to selectively transport their specific signalling molecules while efficiently blocking the passage of all others.

Driven by the notion that natural evolution has optimized the function of such devices, large effort is being currently invested into the creation of artificial nano-molecular sorting devices that mimic the function of biological channels \cite{tijana,caspi-elbaum-2008,martin-antibody-science-2002,martin-DNA-2004,martin-apoenzymes-1997,peters-1999-single-pore,polymer-nanotubes-2008,akin-DNA-nanotubes-2007}. The design of such devices requires detailed understanding of the principles of selective transport through narrow channels.

The precise conditions for the optimal transport selectivity through narrow channels still elude our understanding. A large body of  experimental work indicates that the selectivity is  often based on the differential interactions of the transported molecules with their corresponding transport channels. Moreover, interaction of the transported molecules with their corresponding transport channels is strong, exceeding the interaction of the non-specific competitors. Another salient feature of such channels is that they are narrow, so that the particles cannot freely bypass each other \cite{lim-aebi-review,schulten_glycerol,bezrukov-kullman-maltoporin-2000,bezrukov-antibiotics-PNAS-2002,GlpF-science-2001,martin-DNA-2004,caspi-elbaum-2008,martin-antibody-science-2002,martin-apoenzymes-1997,aquaporins-borgnia,aquaporins-book,mike_review,stewart_review,wente-review,macara-review}.
.

Recent theoretical works have shown that selective transport  through narrow channels can arise from a balance between efficiency and speed; transient trapping inside the channel increases the probability of a molecule to pass through the channel, but leads to jamming at too high trapping times \cite{schulten_glycerol,we-NPC-plos-2007,berezhkovskii_ptr,berezhkovskii-optimal-2005,bezrukov-sites-2005,kolomeisky-2006,schuss-eisenberg-1995,strange-people-occusion-pnas-2006,chou-PRL-single-file-1998,noble-theory-1991,noble-theory-1992,bezrukov-asymmetric-2007}.

Extending previous work, in this paper we have analyzed transport through narrow channels in the framework of generalized kinetic theory. We represent a transport channel as a sequence of positions (sites) and the transport through the channel is determined by the hopping rates from one position to another inside the channel, as well as by the exit ('off') rates from the channel at its ends. To take into account the limited space inside the channel, and the finite size of the transported particles, we allow only a limited occupancy at  each position, $n_m$. Thus, a particle present at a given position along the channel, can hop to an adjacent position only if the latter is occupied by less than $n_m$ particles. Our model allows one to naturally treat channel occupancy by multiple particles and extends the treatment beyond the single-file transport. The main determinant of the transport properties of the channel is  not the interaction strength of the particles with the channel \emph{per se}, but the kinetic properties of the channel, which determine the trapping time $\tau$ and also depend on the geometrical properties of diffusion in the confined space inside the channel. These possibilities are illustrated in the Fig. 2.

We briefly summarize our major findings below. In qualitative agreement with previous work, we find that the transient trapping of the particles in the channel increases the transport probability; particles that have high exit rates do not stay  in the channel long enough to reach the exit into the destination compartment on the right side (cf. Fig. 2), and have a high probability to return back \cite{berezhkovskii_ptr,we-NPC-plos-2007,gardiner-book,schuss-eisenberg-1995,bezrukov-sites-2005,noble-theory-1991,noble-theory-1992}. Essentially, transient trapping increases the time that the particles spend inside the channel to be long enough in order to reach the exit on the right side. Thus, although each individual particle spends more time in the channel, the transmitted flux  is higher. If the only measured quantity is the flux through the channel, experiments can not easily distinguish between the probability of transport and transport speed. \cite{ribbeck-gorlich-EMBO-2001,timney-mike-JCB-2006,bezrukov-kullman-maltoporin-2000,peters-1999-single-pore}. However, they can be distinguished in the experiments that  follow transport of individual molecules \cite{kubitschek_single,musser_single,yang-musser-JCB-2006}.

When the exit rate is too slow or the incoming flux is too high, the rate of particles' entrance to the channel becomes higher than the rate of exit and the particles start to accumulate inside the channel, because the space inside is limited. This leads to two distinct effects. First, the particles inside the channel start to interfere with the passage of  each other. Second, they block the entrance site  and inhibit the entrance of new particles. The channel thus becomes jammed. We must distinguish between translocation \emph{probability} and transport \emph{efficiency}. Transport efficiency is the fraction of the total incoming flux that reaches the exit. Only a certain fraction of the incoming flux can enter the channel because the entrance site can be occupied when particles attempt to enter. This effect decreases the capability to enter the channel and, as a consequence, decreases the transport efficiency. Interactions between the particles inside the channel can also influence the \emph{probability} of individual particles to translocate through the channel upon entering compared to the single particle case. However, we found that for internally uniform channels the crowding of the particles inside the channel \emph{does not} affect the probability of individual particles to translocate through the pore. Thus, the effect of particle accumulation in the channel manifests only in the blocking of the entrance to the channel, which leads to the decrease in the total transport efficiency (and transmitted flux) at low exit rate or high incoming flux (Fig 4\textbf{B}). Thus, we predict that the kinetic profile near the entrance is an important factor in determining the selectivity of transport.

For symmetric channels, this balance between the transport probability and the obstruction of the particle entrance to the channel, determines the optimal exit rate $r_o^{\text{max}} $, (cf. Fig. 4) which maximizes the transport. This provides a basis for selectivity, whereby different molecules can be selected by the kinetics of their transport through the channel \cite{we-NPC-plos-2007,berezhkovskii-optimal-2005,strange-people-occusion-pnas-2006,kolomeisky-2006,chou-PRL-single-file-1998,bezrukov-sites-2005,bezrukov-asymmetric-2007}. In the case discussed in this paper, when many particles can be present in the channel simultaneously, the optimal exit rate and the optimal flux depend on the length of the channel (cf. Section 3). Notably, this is a purely kinetic selectivity mechanism: although a low exit rate can be due to energetic interactions between the transported particles and the channel; the transport efficiency is not determined by the equilibrium occupancy considerations; the selectivity can go beyond the difference in the equilibrium binding affinities between different molecules.

The fact that the transport efficiency has a maximum at a certain value of the exit rate $r_o=r_o^{\text{max}}$ provides a natural definition for the 'jamming transition'. Particles with the exit rates faster than $r_o^{\text{max}} $ pass through the channel essentially unhindered by the interactions with other particles because they do not stay in the channel long enough(Fig. 3\textbf{A}.) On the other hand, particles with exit rates slower than $r_{\text{max}} $ compete with each other for entrance into the limited space inside the channel and the channel becomes jammed. Importantly, we found that the interactions between the particles, and the competition for the limited space inside the channel do not play an important role until quite a few of them accumulate in the channel. For long channels, approximately half of the available channel sites are occupied at the jamming transition ( Fig. 4). This implies that in many experimental situations the interactions between the transported particles do not play a significant role, and may explain why the apparent diffusion coefficient in many flux measurement experiments is found to be almost as high as for free diffusion \cite{ribbeck-gorlich-EMBO-2001,peters-1999-single-pore,mike_review,musser_single}.

Although many particles can be crowded inside the channel, and the entrance to the channel is blocked, transmitted flux  does not disappear even at high fluxes and densities, but rather saturates to the limiting value determined by the trapping time  and the channel length (cf. Fig. 3 and Fig. 5.)
This closely parallels Michaelis-Maenten kinetics of enzymatic reactions \cite{alberts-book,nemenman-sinitsyn} and might be relevant to estimation of binding strengths from flux experiments \cite{kopito-elbaum-PNAS-2007,ribbeck-gorlich-EMBO-2001,timney-mike-JCB-2006}.
Notably,the selective and efficient transport persists beyond the single file transport, even when the ratio of the channel diameter to the particle size is large. In this case, the optimal exit rate $r_o^{\text{max}} $ is simply shifted to lower values.

In order to determine whether the theory developed in this paper can provide an adequate description of experiments, we compared predictions of the theory to the experiments reported in \cite{martin-DNA-2004}. That work found that at low concentrations of the transported particles the flux  through artificial nano-channels increases if the particles can transiently bind inside the channel. Moreover, as the binding energy of the particles was decreased, the enhancement of the flux was lower.  However, as the concentration of the particles in the origin compartment increases, the flux saturates for the channels with transient binding, while the saturation if not observed for non-binding channels, at the experimental range of concentrations. Both these results are in agreement with the theory and can be semi-quantitatively described by the theoretical predictions, as shown in section \ref{section-comparison-with-experiment}.

Thus, we find that the theory based on only two main ingredients: 1) transient trapping of the molecules inside the channel
and 2) crowding of the molecules in the limited space inside the channel, captures the essential features of the selective transport through nano-channels. Moreover, the theory provides verifiable predictions regarding how the flux and selectivity of such channels depend on the channel length, channel radius, the size of the transported molecules and the strength of the interactions of the molecules with the channel.  In particular, we predict that the flux through such channels can be optimized by varying the interaction parameters and the channel dimensions. Such predictions are useful for the design of artificial nano-sorting devices. Further quantitative experiments and comparison with the theory are needed in order to test the theory and for its further refinement.

We expect that the effects described in this paper should play a role in selective transport through any narrow channel.
For instance, the effects described in this paper might be relevant in determining the selectivity of the ion channels, although other factors  might be dominant \cite{hille-book,jordan-review,corry-chung-valence-selectivity-2006,eisenberg-barriers-2007}.
In each particular system other effects related to molecular details might be dominant determinants of selectivity. Such effects might include the long range electrostatics and channel fluctuations in the ion channels, the details of the transfer of the transported molecules from one binding moiety to another, and conformational changes of the filaments that carry the binding moieties (as in the nuclear pore complex and other polymer-based systems).

Finally, we note that the theory developed in this paper can also be applied to other signal-transducing schemes, such as signalling cascades and multi-step enzymatic reactions \cite{enzyme-ACH-pnas-1998,ramanathan-broach-cross-talk-2007,mckeithan-proofreading-pnas-1995,nemenman-sinitsyn}.

\begin{small}
The author is thankful to C. Connaughton, B. Chait, I. Nemenman, J. Pearson, A. Perelson, Y. Rabin, K. Rasmussen, M. Rout, N. Sinitsyn, T. Talisman, Z. Schuss for stimulating discussions, P. Welch for comments on the manuscript, and  anonymous reviewers for helpful suggestions. This research was performed under the auspices of the U.S. Department of Energy under contract DE-AC52-06NA25396.
\end{small}


\section*{Appendix}
\subsection*{Single particle occupancy: connection to previous work}
In this section we show that the  model of this paper can be reduced to previous models, in a proper limit. Let us assume, following \cite{bezrukov-sites-2005,bezrukov-asymmetric-2007,berezhkovskii-optimal-2005,kolomeisky-2006,strange-people-occusion-pnas-2006} that already when the channel is occupied only by one particle, it prevents the entrance of others. The channel, however, is long, and the particle can obey complicated kinetics inside, which determines its probability to traverse the channel, and the time it spends inside. Physically, such situation can arise, for instance, due to strong long-range repulsion between the particles.

In this case, the problem reduces to a 'single-site' channel of Sec. \ref{section-single-site} but with forward exit rate $r_{\rightarrow} $, backward exit rate $r_{\leftarrow}$ that are not independent, but are determined by the internal kinetics of the channel, and are related through the single-particle dwelling time $\tau$ and transport probability $P_{\text{tr}}$.  As in Sec. \ref{section-single-site}, the transmitted flux is
\begin{equation}
J_{\text{out}}=\frac{Jr_{\rightarrow}}{J+r_{\rightarrow}+r_{\leftarrow}}
\end{equation}

From equation (\ref{Jout_with_exclusion}), the probability of a single particle to traverse the channel of length $N$ (for $J\rightarrow 0$) is $P_{\text{tr}}=1/(2+(N-1)N/(\tau r))=r_{\rightarrow}/(r_{\rightarrow}+r_{\leftarrow})$, and the residence time is $\tau=N/(2r_o)=1/(r_{\rightarrow}+r_{\leftarrow})$ \cite{pearson-BJ-2005}. Thus, we get for the transmitted flux:
\begin{equation}
J_{\text{out}}=\frac{J}{2\left(1+J\tau\right)\left(1+\frac{(N-1)N}{2\tau r}\right)}
\end{equation}
which is identical to  expressions obtained in Ref.\cite{bezrukov-sites-2005}, if one bears in mind that  the flux is $J=k_{\text{on}}c$, where $c$ is the concentration of the particles outside the channel.

In is important to note that the optimal exit rate in this case is $r_o^{\text{max}}=\sqrt{\frac{JrN}{N-1}}$, that is almost independent of $N$ for long channels. This is in contrast to the model of  Sec. 2.4, which takes into account multiple occupancy of the channel by many particles - where the optimal exit rate decreases with $N$. The optimal current is, by contrast, higher for multiple-occupance channels. This is natural - if more particles are can occupy the channel before it becomes jammed, the channel can sustain a higher current.
\subsection*{Connection between continuum and discrete models.}
Discrete model of equation (\ref{kinetics_with_exclusion}) reduces to a continuum description of transport inside the channel, if one defines the one-dimensional particle density $c^1(x)=n_i/a$ where $a$ is the distance between the 'sites', so that $x=ai$, with a diffusion coefficient $D_{\text{in}}=a^2r$ \cite{bezrukov-sites-2005,gardiner-book,berg-book}. For comparison with real systems, one-dimensional diffusion inside the channel must be matched to the three-dimensional diffusion outside the channel, through the choice of $r_o$  (see e.g.\cite{berg-book,peters-1999-single-pore,bezrukov-sites-2005,bezrukov-3d-2000}). For clarity, we re-derive this connection here without the inter-particle interactions inside the channel - see Fig. 6 for illustration.

We denote the three-dimensional concentration of particles at the left side far away from the channel as $c_L^{\infty}$; we assume that concentration on the right side far away from the channel is zero. At steady state, a density profile will be established such that the flux through the pore is $F$, the (three-dimensional) density at the pore entrance on the left is $c_L$ and the density at the exit on the right is $c_R$; the corresponding one-dimensional densities are $c^1_L=c_L \beta R^2$, and $c^1_R=c_R \beta R^2$, where $R$ is the channel radius, and $\beta $ is a geometrical pre-factor that depends on the shape of the channel opening ($\beta =\pi$ for circular opening).

At steady state, the flux that enters  the channel from the left is \cite{berg-book}:
\begin{eqnarray}
J=\alpha (c^{\infty}_L-c_L)RD_{\text{out}}=F
\end{eqnarray}
where $\alpha $ is a geometrical pre-factor that depends on the shape of the channel opening; $\alpha=4$ for a circular opening\cite{berg-book}. Note that if all the impinging particles would go through the channel, the entering flux would be $J_0=\alpha c^{\infty}_LRD_{\text{out}}$ - the flux to a fully absorbing patch of radius $R$\cite{berg-book}. However, even in the absence of jamming, not all particles go through - some of them return back, after hopping back and forth inside the channel, as reflected in the returned portion of the flux $-\alpha c_L R D_{\text{out}}$\cite{berg-book}.

The flux that exits the channel to the right is \cite{berg-book,bezrukov-sites-2005}:
\begin{eqnarray}
J_{\text{out}}=\alpha c_RRD_{\text{out}}=F
\end{eqnarray}
The flux inside the channel, for a flat potential profile, is: \cite{we-NPC-plos-2007,bezrukov-3d-2000,gardiner-book,berezhkovskii_ptr}
\begin{eqnarray}
F=\frac{c^1_L-c^1_R}{L Z}D_{\text{in}}
\end{eqnarray}
where $Z=\langle e^E\rangle$ is the average inverse Boltzmann factor of the attractive energy inside the channel, $E<0$.
Solving the above equations, we get:
\begin{eqnarray}
F=\frac{J_0}{2+\frac{\alpha}{\beta}\frac{L}{R}\frac{D_{\text{out}}}{D_{\text{in}}}Z}
\end{eqnarray}
And thus the fraction of the transmitted flux is
\begin{equation}
P_{\text{tr}}=\frac{1}{2+\frac{\alpha}{\beta}\frac{L}{R}\frac{D_{\text{out}}}{D_{\text{in}}}Z}
\end{equation}
On the other hand, equation (\ref{Jout_with_exclusion}) gives without jamming ($J\rightarrow 0$)
\begin{equation}\label{q}
P_{\text{tr}}=\frac{1}{2+(N-1)r_o/r}=\frac{1}{2+\frac{L}{a}\frac{r_o}{r}}
\end{equation}

Finally, choosing $r_o/r=\frac{J_{\text{out}}Z/n_N}{a^2/D_{\text{in}}}=\frac{\alpha}{\beta}\frac{D_{\text{o}}}{D_{\text{in}}}\frac{a}{R}Z$, the discrete and continuous formulations become equivalent as long as $N=L/a\gg 1$ \cite{bezrukov-3d-2000,bezrukov-sites-2005,gardiner-book,chou-tasep-2003}.

The distance between the 'sites' models the excluded volume interactions between the particles. In this paper we make the most parsimonious choice: the distance between  sites is equal to the size of the particle. This choice adequately captures the essential properties of hindered diffusion in narrow channels \cite{berezhkovskii-hummer,chou-PRL-single-file-1998,chou-zeolites-PRL-1999edp,chou-tasep-2003,schulten_glycerol}. In principle, in some systems the actual 'diffusion step' can be smaller than the particle size. However, in known cases, the results remain qualitatively the same after proper re-scaling of the transition rates \cite{zia-shaw-arb-size-2003,shutz-arb-size-2004,chou-arb-size-2003,macdonald-gibbs-1968,schulten_glycerol}. We also found that the quality of the fits in Fig. \textbf{5} and overall conclusions are not sensitive to small variations in the estimates of the parameters of the model (data not shown).

\subsection*{Expressions for $M\neq 1$}
For completeness, we present here the expressions for the general case $1\leq M< N/2$, $n_m=1$
The optimal exit rate (for the values of $J,M$and $N$ when the optimum exists):
\begin{eqnarray}
r^{\text{max}}_o/r=\frac{J/r (1-M)-\sqrt{J/r (-2 M+N+1)}}{J/r (M-1)^2+(2 M-N-1)}
\end{eqnarray}
The  channel occupancy
\begin{eqnarray}
\frac{J/r (2+(N-1)r_o/r)(N+(M-1)(N-M)r_o/r)}{2(r_o/r(2+(N-1)r_o/r)+J/r(1+(M-1)r_o/r) (1+(N-M)r_o/r))}
\end{eqnarray}
and the saturation current in the $J/r\rightarrow\infty$ limit:
\begin{eqnarray}
\frac{r_o(1+(M-1)r_o/r)}{(M-1)(N-M)(r_o/r)^2+(N-1)r_o/r+1}
\end{eqnarray}

\clearpage
\section*{Figure Legends}
\subsubsection*{Figure~1.}
\textbf{Schematic diagram of transport through a channel}
\textbf{A.} Schematic illustration of the transport through a narrow channel. \textbf{B.} Kinetic diagram of a one-site channel. \textbf{C.} Kinetic diagram of a two-site channel.

\subsubsection*{Figure~2.}
\textbf{Kinetic diagrams of transport through a channel of an arbitrary length}
\textbf{A.} Symmetric channel consisting of $N$ positions (sites). The  particles enter the channel at a site $M$ with an average rate $J$. \textbf{B.} Equivalent energetic diagram in the case when the exit rates are determined by the interaction (binding energy) with the channel. The exit rates at the channel ends are given by Arrhenius-Boltzmann  factors of the energy barriers at the exits, $E_{\rightarrow}$ and $E_{\leftarrow}$: $r_{\rightarrow}\sim \exp(-E_{\rightarrow}/kT)$ and $r_{\leftarrow}\sim \exp(-E_{\leftarrow}/kT)$
\textbf{C.} Equivalent geometry of the channel in the case when the exit rates are due to spatial bottlenecks at the channel ends.

\subsubsection*{Figure~3.}
\textbf{Efficiency of transport through a channel of an arbitrary length}
\textbf{A.} Transport efficiency as a function of the exit rate for $J/r=0.01 $, $n_m=1$ for different entrance sites $M$. Black line: $M=1, N=10$, gray line: $M=4, N=40 $; corresponding dashed lines show the \emph{probability} of a particle to traverse the channel; it is identical to a single particle transport efficiency in the limit $J\rightarrow 0 $ - cf. text. \textbf{B.} Transport efficiency as a function of channel length $N$, for the optimal value of exit rate $r_o=(Jr/(N-1))^{1/2}$, $M=1$, $J/r=0.01 $, $n_m=1$.  Black line: $J/r=0.01$, dashed line $J/r=0.1$ \textbf{C.} Transmitted flux $J_{\text{out}}/J_{\text{out}}^{\infty} $ - cf. equations (8) and (16), as a function of the normalized incoming flux $J/r$; black line: $r_o/r=0.01$, dashed line $r_o/r=1$; $M=1$;$n_m=1$. Note that the transmitted flux saturates to a constant value $J_{\text{out}}^{\infty}$ in the jammed regime. \textbf{D.} Optimal exit rate as a function of the channel length $N$ for $M=1$, $J/r=0.01$, $n_m=1$, (black line). Dashed line: same for $J/r=0.1$.

\subsubsection*{Figure~4.}
\textbf{Occupancy of the channel at the jamming transition} \textbf{A.} Occupied fraction  of the channel at the jamming transition, $r_o=r_{\text{max}}$, as a function of the  channel length $N$, for different values of the incoming flux $J/r$. It shows that the channel can be occupied to a considerable degree - up to half of the available sites - before the jamming becomes significant. \textbf{B.}  Densities at the entrance site $1$ (black line) and exit site $N$ (dashed line) as a function of the incoming flux $J/r$ for $r_o/r=0.1$, $N=5$, $n_m=1$. Density at the entrance site saturates to $1$, which causes the saturation of the transmitted flux. Density at the exit site stays low even in the regime when the transmitted flux through the pore saturates.

\subsubsection*{Figure~5.}
\textbf{Flux through nano-channels: comparison with experiment} \textbf{A.} Flux through the nano-channel as a function of the outside concentration of the transported ssDNA. Black dots - experimental data from Ref.\cite{martin-DNA-2004} for a nano-channel without trapping inside. Corresponding black line - theoretical fit from eq. (\ref{Jout_with_exclusion}) with $n_m=6$, Z=1, $D_{\text{in}}/D_{\text{out}}=0.42$, $N=L/(2S)$. Red dots - experimental data from Ref.\cite{martin-DNA-2004} for a nano-channel with ssDNA hairpins grafted inside the channel, which are complementary to the transported ssDNA. Corresponding red line is the theoretical prediction of eq. (\ref{Jout_with_exclusion}) with $n_m=3$, $D_{\text{in}}/D_{\text{out}}=0.0042$, $Z=0.00007$, $N=L/(2S)$  \textbf{B.} Reduction of the flux through the channel as a function of the number of mismatches between transported ssDNA and the ssDNA hairpins grafted inside, relative to the flux of the perfect complement ssDNA measured at the feed ssDNA concentration $9\mu$M. Dots - experimental data from Ref.\cite{martin-DNA-2004} for a single mismatch at the edge of the transported DNA segment; square - single mismatch in the middle of the transported ssDNA segment; line - theoretical model; same parameter values as used in panel \textbf{A} - cf. text.

\subsubsection*{Figure~6.}
\textbf{Three-dimensional diffusion outside the channel} Schematic illustration of the three-dimensional diffusion outside the channel and one-dimensional diffusion inside. See text in Appendix.

\clearpage

\begin{figure}[htbp]\label{fig-schematic-one-two}
\centerline{
\includegraphics[width= 10 cm]{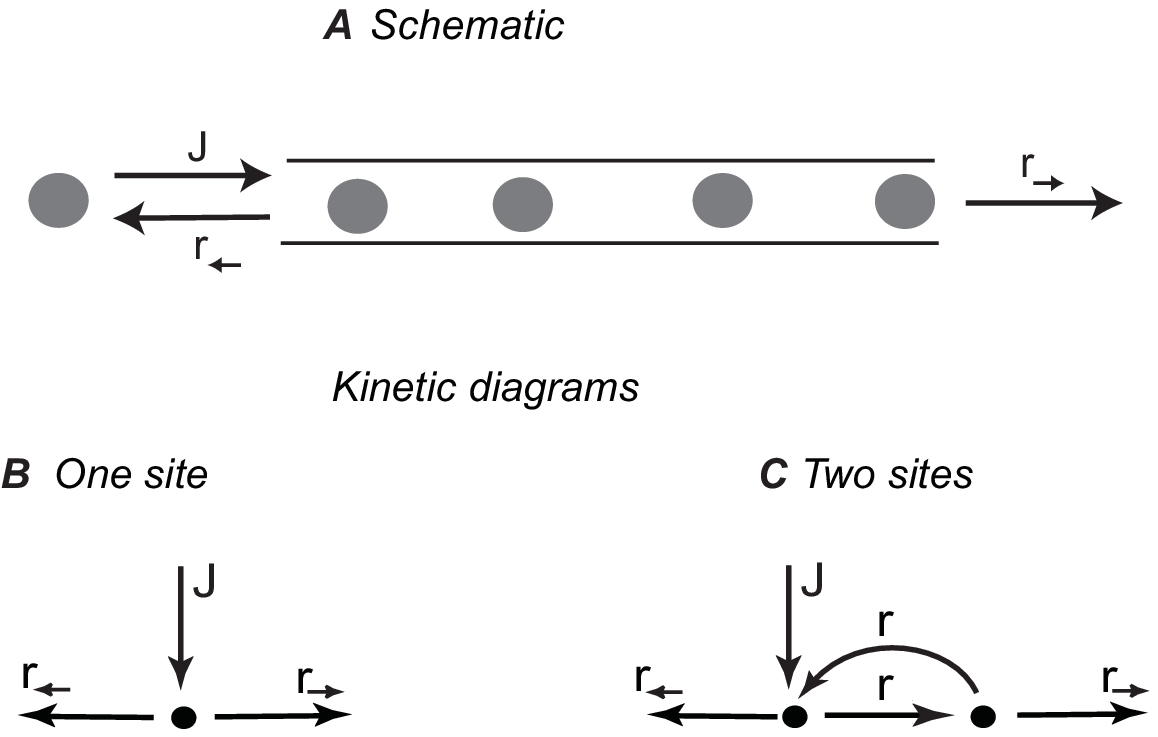}
}
\caption{}
\end{figure}

\clearpage

\begin{figure}[htbp]\label{fig-N-site-kinetics}
\centerline{
\includegraphics[width= 12 cm]{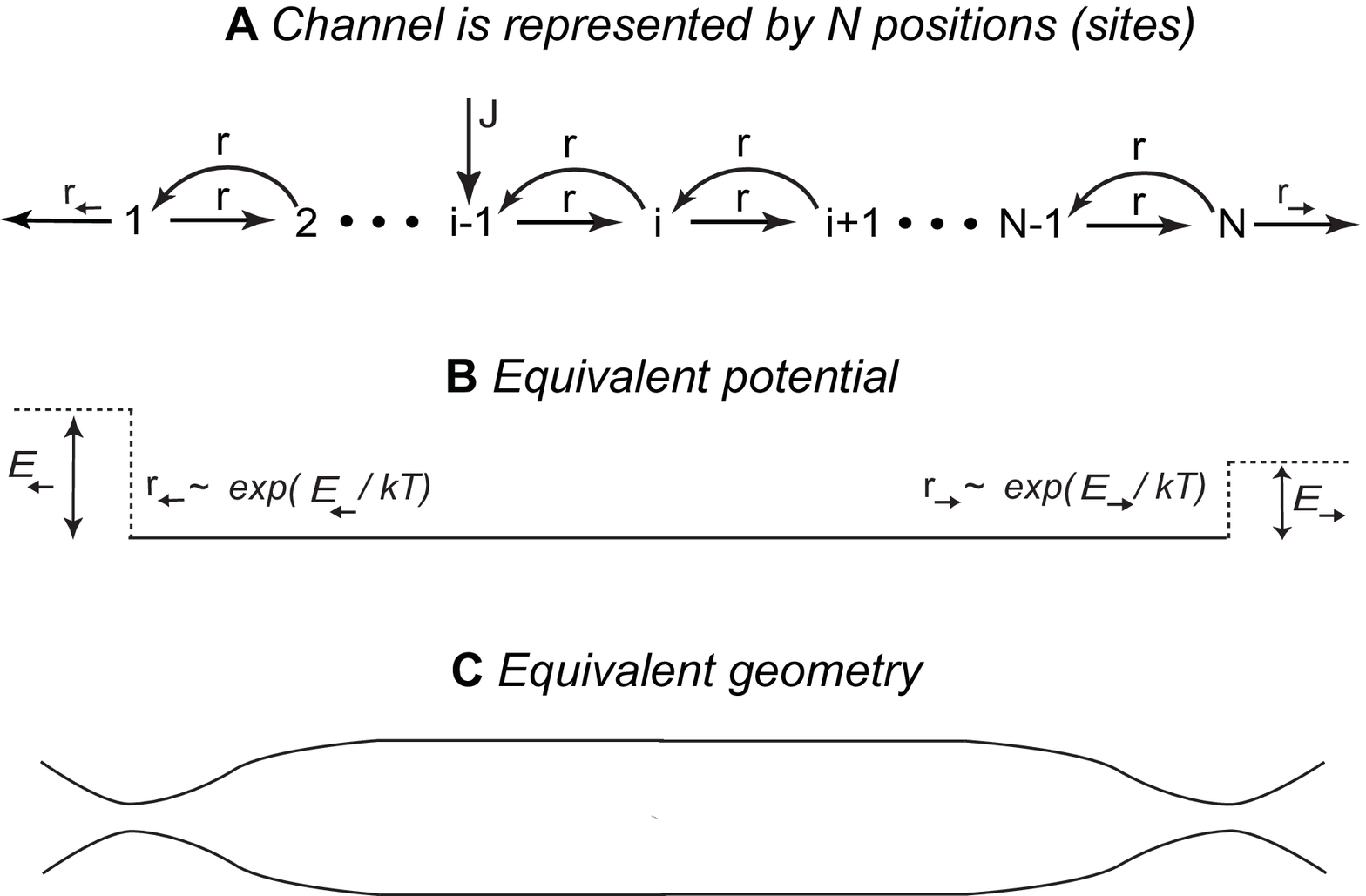}
}
\caption{}
\end{figure}

\clearpage

\begin{figure}[htbp]\label{fig-selectivity-nocomp}
\centerline{
\includegraphics[width= 14 cm]{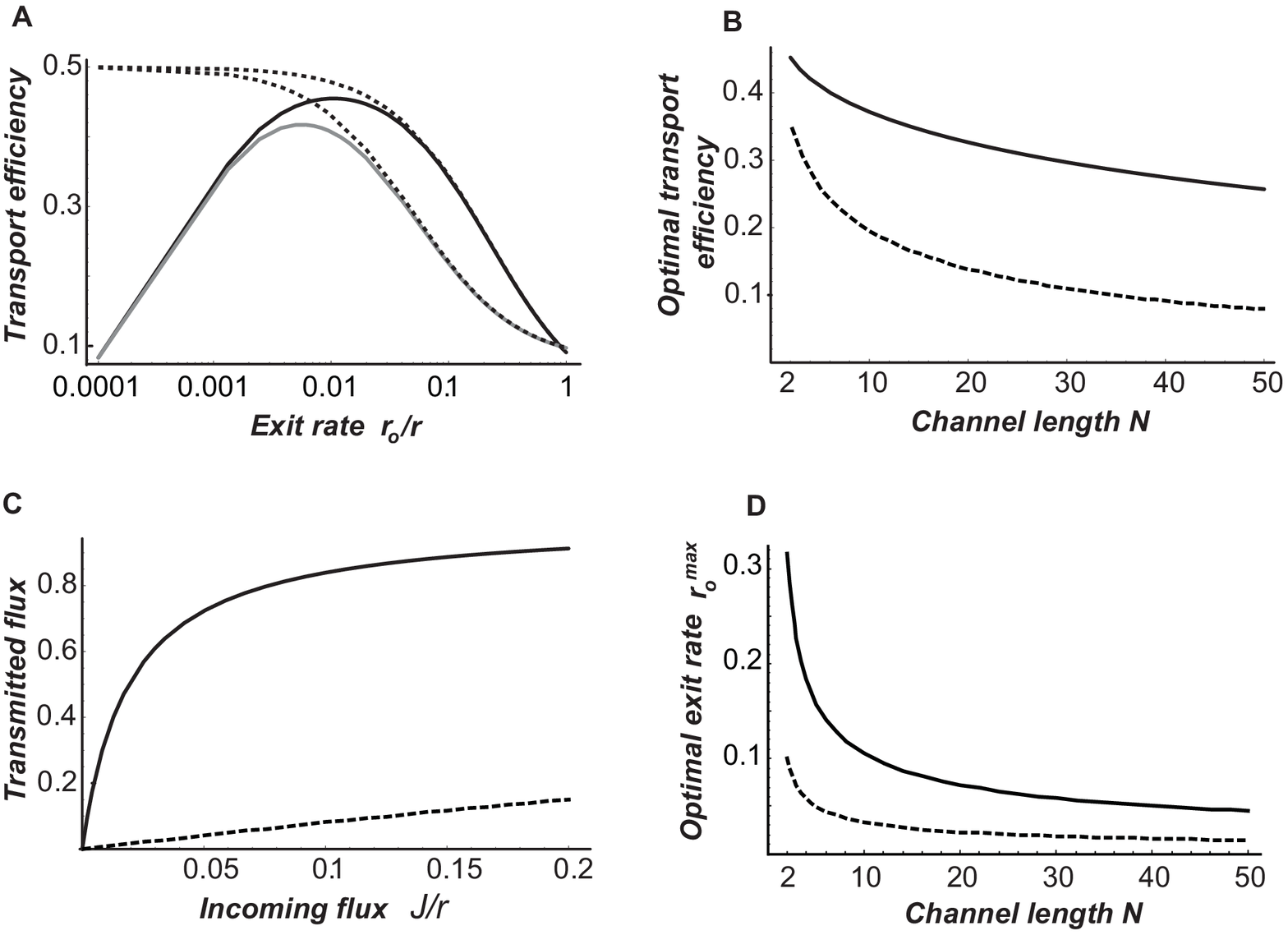}
}
\caption{}
\end{figure}

\clearpage

\begin{figure}[htbp]\label{fig-occupancy-at-jamming}
\centerline{
\includegraphics[width= 14 cm]{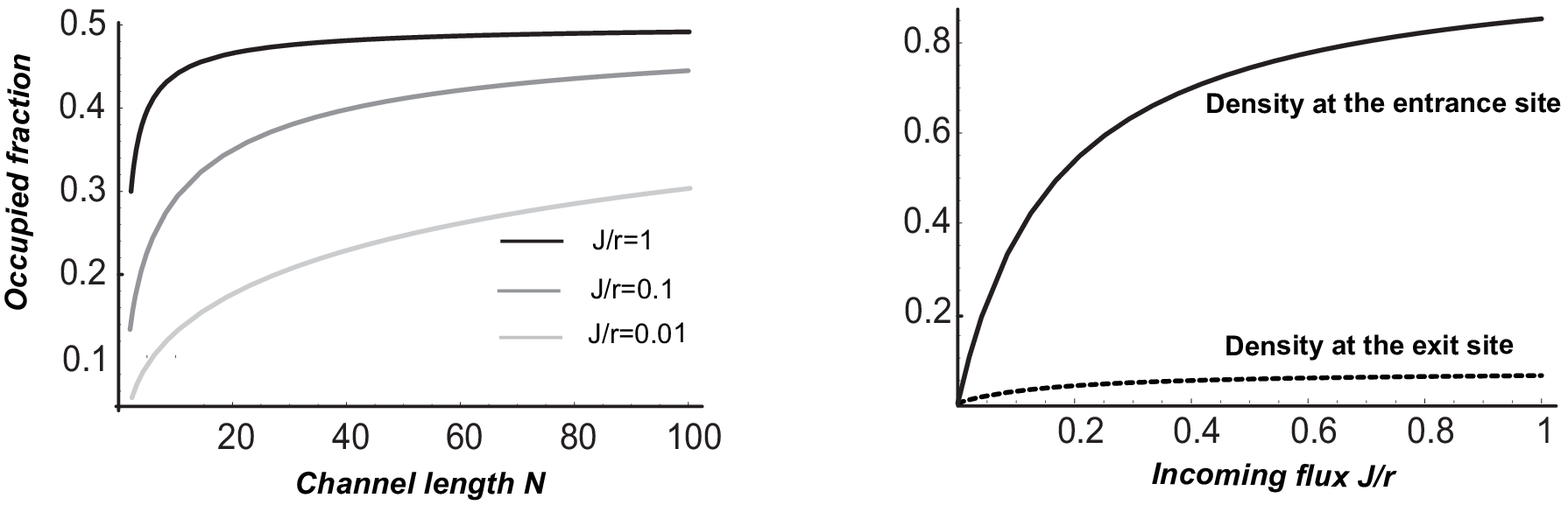}
}
\caption{}
\end{figure}

\clearpage

\begin{figure}[htbp]\label{comparison-with-experiment}
\centerline{
\includegraphics[width= 14 cm]{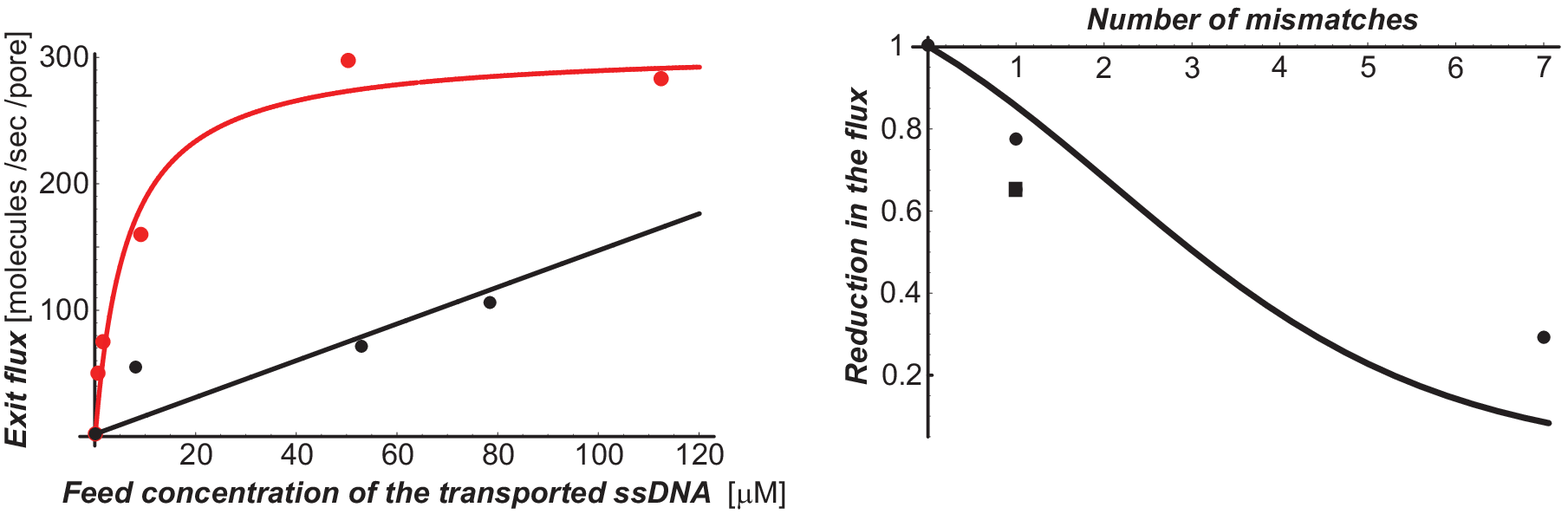}
}
\caption{}
\end{figure}

\begin{figure}[htbp]\label{fig-appendix-3d}
\centerline{
\includegraphics[width= 10 cm]{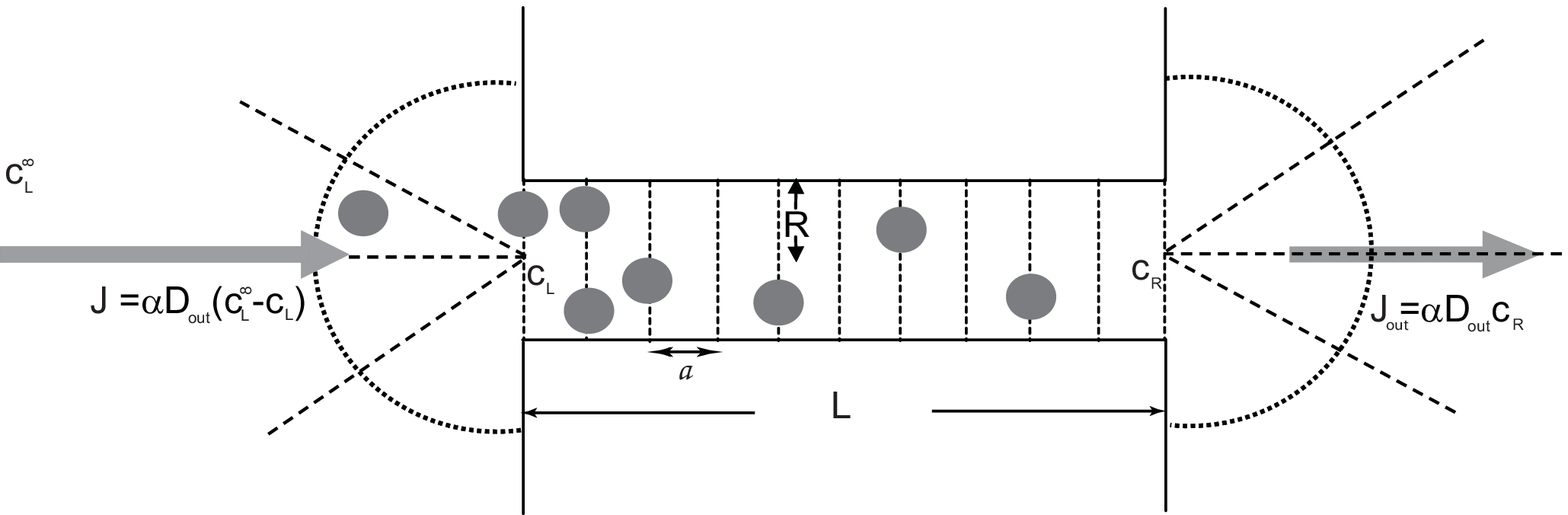}
}
\caption{}
\end{figure}

\end{document}